\documentclass[aps,pra,showpacs,preprint,superscriptaddress,floatfix,longbibliography]{revtex4-1}
\usepackage[latin1]{inputenc}
\usepackage[english]{babel}
\usepackage[T1]{fontenc}

\usepackage{graphicx}
\usepackage{lastpage}
\usepackage{amsmath,amssymb}
\usepackage{hyperref}

\usepackage{braket}
\newcommand{\dpa}[2]{\frac{\partial #1}{\partial #2}} 
\newcommand{\dpat}[1]{\frac{\partial^2 }{\partial {#1} ^2}} 
\newcommand{\dd}[1]{\frac{ d^2 }{d {#1} ^2}} 

\newcommand{\fed}[1]{\mathbf{#1}}

\newcommand{\abs}[1]{\left| #1 \right|}

\newcommand{\Vcomma}{;}
\newcommand{\qx}{q}
\newcommand{\elecvar}{z}

\DeclareMathOperator{\imag}{Im}
\DeclareMathOperator{\real}{Re}

\usepackage{siunitx}
\usepackage[draft]{fixme}
\begin{document}
\title{Theory of dissociative tunneling ionization}

\author{Jens Svensmark}
\affiliation{Department of Physics and Astronomy, Aarhus University, 8000 Aarhus C, Denmark}
\author{Oleg~I.~Tolstikhin}
\affiliation{Moscow Institute of Physics and Technology,
Dolgoprudny 141700, Russia}
\author{Lars Bojer Madsen}
\affiliation{Department of Physics and Astronomy, Aarhus University, 8000 Aarhus C, Denmark}
\date{\today}

\begin{abstract}

We present a theoretical study of the dissociative tunneling ionization process. Analytic expressions for the nuclear kinetic energy distribution of the ionization rates are derived. A particularly simple expression for the spectrum is found by using the Born-Oppenheimer (BO) approximation in conjunction with the reflection principle. These spectra are compared to exact non-BO \emph{ab initio} spectra obtained through model calculations with a quantum mechanical treatment of both the electronic and nuclear degrees freedom. In the regime where the BO approximation is applicable imaging of the BO nuclear wave function is demonstrated to be possible through reverse use of the reflection principle, when accounting appropriately for the electronic ionization rate. A qualitative difference between the exact and BO wave functions in the asymptotic region of large  electronic distances is shown. Additionally the behavior of the wave function across the turning line is seen to be reminiscent of light refraction. For weak fields, where the BO approximation does not apply, the weak-field asymptotic theory describes the spectrum accurately.

\end{abstract}


\maketitle


\section{Introduction}

Currently a number of intense mid-infrared light sources are being developed \cite{Popmintchev1287,Silva15,PhysRevX.5.021034,Duval15}, spurred on by their uses in sub-attosecond pulse generation \cite{PhysRevLett.98.013901,PhysRevLett.111.033002}, strong-field holography \cite{Huismans61,PhysRevA.84.043420} and laser-induced electron diffraction \cite{PhysRevX.5.021034,Blaga12}. The low frequency and high intensity of these new sources mean that the tunneling picture is an appropriate framework for describing how these light sources interact with atoms and molecules. In this work we deal with the process of dissociative tunneling ionization in molecules, where a static electric field tunnel ionizes an electron, after which the nuclei dissociate. To our knowledge this is the first work on the theory of dissociative tunneling ionization. In the theory we treat the nuclear and electronic degrees freedom on an equal footing and fully quantum mechanically.

The reflection principle~\cite{PhysRev.32.858,:/content/aip/journal/jcp/58/9/10.1063/1.1679721,PhysRevLett.108.073202,PhysRevA.90.063408} is often used to describe the process of dissociative ionization. This principle can be applied within the framework of the Born-Oppenheimer (BO) approximation to relate the nuclear kinetic energy release (KER) spectrum to the nuclear wave function. It was formulated as early as 1928~\cite{PhysRev.32.858}, and later put on a more rigorous foundation~\cite{:/content/aip/journal/jcp/58/9/10.1063/1.1679721} (see also references therein for a list of early uses). 
In time-dependent cases where the time-scale of the electric field is shorter than that of nuclear motion, one assumes that the electrons make an instantaneous Frank-Condon transition to a dissociative electronic state. The probability distribution in the new electronic state is then the absolute value squared of the initial nuclear wave function times some dipole coupling factor. In case this dipole coupling factor is almost constant, the wave packet that enters the dissociative state is practically identical to the initial nuclear wave function. Classical energy conservation then dictates that the nuclear KER spectrum can be obtained by reflecting the nuclear wave function in the dissociative potential curve. This is the regime considered mostly in the literature
. In the time-independent tunneling case, the electronic ionization rate takes the same role as the dipole coupling factor in the time-dependent case, that is, it multiplies the nuclear wave function before it enters the continuum. However, this electronic rate has an exponential dependence on the internuclear coordinate and can by no means be considered constant (as was also pointed out in Ref.~\cite{PhysRevLett.92.163004}). It is therefore essential to consider the effect of this additional factor on the KER spectrum; the spectrum cannot be found simply by reflection of the nuclear wave function.

Imaging of the nuclear wave function is made possible through the reflection principle, by applying it in reverse on a measured KER spectrum \cite{PhysRevLett.82.3416,PhysRevLett.108.073202,PhysRevA.58.426}. This is often referred to as Coulomb explosion imaging. In the tunneling case the exponential dependence of the electronic rate on the internuclear coordinate means that the product of the electronic rate and the nuclear wave function is essentially different from the bare nuclear wave function, and the electronic rate therefore needs to be included to image the nuclear wave function based on the KER spectrum. In Ref.~\cite{rtd} it was demonstrated that the BO approximation breaks down for weak fields. In this case the weak-field asymptotic theory (WFAT) \cite{PhysRevA.84.053423,PhysRevA.89.013421} provides us with accurate results for the KER spectrum.

The paper is organized as follows. In Sec.~\ref{sec:theory} the theory for dissociative tunneling ionization of homonuclear molecules is outlined. We derive an exact expression for the KER spectrum and a corresponding expression in the BO approximation. Section~\ref{sec:1d-calculation} exemplifies the theory with numerical reduced dimensionality calculations. Numerically exact KER spectra are compared to KER spectra obtained in the BO approximation using the reflection principle. Imaging of the nuclear part of the wave function from the KER spectrum is demonstrated. Section~\ref{sec:conclusion-outlook} concludes the paper. Atomic units $\hbar=m_e=e=1$ are used throughout.

\section{Theory}
\label{sec:theory}

\subsection{Basic equations}

We consider a three-body system consisting of two heavy nuclei with masses $m_1=m_2$ and charges $q_1=q_2$, and one electron with mass $m_3=1$ and charge $q_3=-1$. In the center-of-mass frame these have coordinates $\fed{r}_1,\fed{r}_2$ and $\fed{r}_3$ fulfilling $m_1(\fed{r}_1 +\fed{r}_2)+\fed{r}_3=0$. Let us introduce the reduced masses
\begin{align}
  M & = \frac{m_1}{2}, &
  m &  = \frac{2m_1}{2m_1+1} , 
\end{align}
effective charge
\begin{align}
  \qx & = \left(\frac{q_1}{m_1}+1\right) m, 
\end{align}
and Jacobi coordinates 
\begin{align}
 \fed{R}&=\fed{r}_2-\fed{r}_1,& \fed{r}&=\frac{\fed{r}_3}{m}.
\end{align}
We assume that the orientation of the internuclear axis $\fed{R}$ is fixed in space. 
We also assume that the field is directed along the $z$-direction, $\fed{F}=F\hat{\fed{z}}$ and choose to consider $F\geq 0$ for definiteness. Due to the azimuthal symmetry of the molecule only the polar angle $\beta$ between $\fed{R}$ and $\fed{F}$ matters. This $\beta$ angle takes the role as an external parameter, and we omit explicit reference to it in the following. With these assumptions we can write the time-independent Schr\"odinger equation (SE) within the single-active-electron approximation as
\begin{align}
 \left[- \frac{1}{2M} \dpat{R}  - \frac{1}{2m} \Delta_{\fed{r}} +  U(R) + V\left(\fed{r}\Vcomma R\right)  +   \qx zF - E(F)\right] \Psi(\fed{r},R) & = 0,\label{eq:schrodinger}
\end{align}
where the effective $U(R)$ potential describes how the nuclei interact with each other and the effective $V(\fed{r}\Vcomma R)$ potential describes how the electron interacts with the nuclei. For a system with several electrons the $U(R)$ potential represents the BO potential of the system without the active electron. In this work, we assume that $U(R)$ is monotonically decreasing, i.e., it corresponds to a purely dissociative BO curve. 

We assume that the nuclei cannot pass through each other. This gives the boundary condition
\begin{align}
  \Psi(\fed{r},R=0)=0,\label{eq:zero_bc}
\end{align}
and we consider Eq.~(\ref{eq:schrodinger}) in the interval $0\leq R$.
We also impose outgoing-wave boundary conditions in the electronic coordinate $\fed{r}$, the exact form of these will be specified below. With these boundary conditions the wave function we seek as a solution of Eq.~(\ref{eq:schrodinger}) is a Siegert state \cite{PhysRev.56.750,PhysRevLett.79.2026,starka}, with a complex energy $E(F)=\mathcal{E}(F)-\frac{i}{2}\Gamma(F)$, where $\Gamma(F)$ is the ionization rate, and it is normalized by
\begin{align}
  \int d^3\fed{r} \int_0^\infty dR\ \Psi^2(\fed{r},R) & = 1.
\end{align}
The outgoing-wave boundary condition in the electronic coordinate means that the solution we seek to Eq.~(\ref{eq:schrodinger}) describes tunneling of the electron. This tunneling is followed by dissociation of the nuclei for the considered class of strictly dissociative potentials $U(R)$. In the following we will describe the energy distribution of the dissociated nuclei.

\subsection{Energy distribution of dissociated nuclei}
\label{sec:asymptotics}
Our aim is to describe the energy distribution of the nuclei, i.e., the KER spectrum, after the molecule is ionized by tunneling of the electron. To this end we need to consider the problem in the $r\to\infty$ limit, where the electron is far away from the nuclei. In this limit we assume that the electron-nuclear interaction potential takes the form
\begin{align}
  V(\fed{r}\Vcomma {R})|_{r\to\infty} & = -\frac{Z}{r}+O(r^{-2}),
\end{align}
where $Z=2q_1$ is the total charge of the remaining core system. 
This assumption makes our problem separable in electron and nuclear coordinates in this asymptotic region. By seeking the partial solutions in the form $\Psi(\fed{r},R)=f(\fed{r},k)g(R,k)$, Eq.~(\ref{eq:schrodinger}) can be written as the separated equations
\begin{subequations}
  \begin{align}
    \left[ - \frac{1}{2m} \Delta_{\fed{r}}  -\frac{Z}{r}  +   \qx Fz  - E_{\fed{r}}\right] f(\fed{r},k) & = 0,\label{eq:as_x_eq}\\
    \left[- \frac{1}{2M} \dd{R}   +  {U}(R)   - E_{R}\right] g(R,k) & = 0,\label{eq:as_R_eq}
  \end{align}
\end{subequations}
with separation constants given by
\begin{subequations}
\begin{align}
  E(F) & = E_{\fed{r}} + E_{R},\\
  E_{R} & = \frac{k^2}{2M},
\end{align}
\end{subequations}
where we assume $U(R)|_{R\to\infty}=0$ and $k\geq 0$ is the wave number for the state $g(R,k)$.
Equation~(\ref{eq:zero_bc}) amounts to
\begin{align}
    g(R=0,k) & = 0.\label{eq:zero_bc_g}
\end{align}
We choose the continuum solutions of Eq.~(\ref{eq:as_R_eq}) to be real and normalized by
\begin{align}
  \int_0^\infty g(R,k)g(R,k') dR & = 2\pi\delta(k-k').\label{eq:g_norm}
\end{align}
The conditions Eqs.~(\ref{eq:zero_bc_g})-(\ref{eq:g_norm}) completely specify the nuclear problem Eq.~(\ref{eq:as_R_eq}). 

The electronic problem Eq.~(\ref{eq:as_x_eq}) has a potential consisting of a Coulomb term and a linear field term. This problem is separable in parabolic coordinates~\cite{landau1977quantum}, which we will therefore use.
First we introduce mass-scaled quantities
\begin{subequations}
\begin{align}
  \tilde{\fed{r}} & = \sqrt{m} \fed{r},\\
  \tilde{F} & = \frac{q}{\sqrt{m}}F\\
  \tilde{Z} & = \sqrt{m}Z.
\end{align}
\end{subequations}
Then the following form of the parabolic coordinates is introduced (as in Ref.~\cite{PhysRevA.84.053423})
\begin{subequations}
\label{eq:parab_coord}
  \begin{align}
    \xi &  = \tilde{r} + \tilde{z}, &0\leq \xi <\infty\\
    \eta &  =  \tilde{r} -  \tilde{z}, &0\leq \eta <\infty\\
    \varphi &  = \arctan\frac{ \tilde{y}}{ \tilde{x}}, &0\leq \varphi <2\pi.
  \end{align}
\end{subequations}
With this choice of coordinates a potential barrier forms in the $\eta$ coordinates and therefore $\eta$ takes the role as the 'tunneling coordinate'.

\begin{figure}
  \centering
  \includegraphics{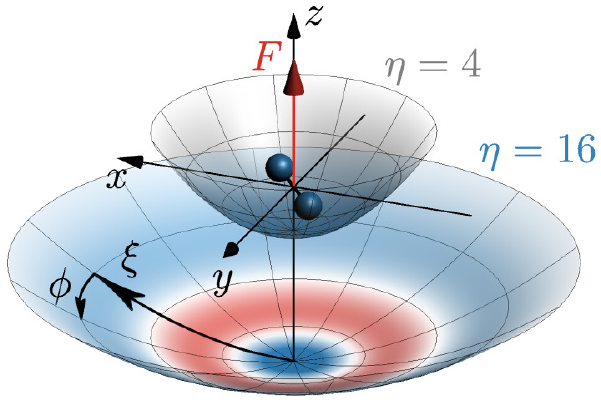}
  \caption{(Color online) The blue/red surface is a paraboloid showing a surface of constant $\eta$. The gray paraboloid is the same for a smaller value of $\eta$. The electron is ionized in the negative $z$ direction due to its negative charge, given that the electric field points in the positive $z$-direction. The $\Phi_\nu(\xi,\varphi)$ states [Eq.~(\ref{eq:xi_phi_ad_eq})] 'live' in the constant $\eta$ paraboloids. The colors in the blue/red surface illustrates an example of the nodal structure of such a $\Phi_\nu(\xi,\varphi)$ state. The curvature of the paraboloids means that the $\Phi_\nu(\xi,\varphi)$ states are bound. This means that $\eta$ is the only coordinate where we have to consider the wave function at infinity, i.e., $\eta$ is the 'tunneling' coordinate. For large $\eta$ the polar angle $\beta$, which specifies the orientation of the molecule, does not matter for the asymptotic form of the wave function in parabolic coordinates, though it matters for the size of the coefficients [Eq.~(\ref{eq:spectrum_ampl_def})].}
\label{fig:parab_coord}
\end{figure}

In the asymptotic region $\eta\to\infty$ Eq.~(\ref{eq:as_x_eq}) has a solution that is a linear combination of partial solutions of the form \cite{PhysRevA.84.053423}
\begin{align}
  f(\fed{r},k)|_{\eta\to\infty} & =  \eta^{-1/2}f(\eta)\Phi_\nu(\xi,\varphi),
\end{align}
where the outgoing-wave $f(\eta)$ is given by
\begin{align}
  f(\eta) &= \frac{2^{1/2}}{(\tilde{F}\eta)^{1/4}}\exp\left(i\left[\frac{\tilde{F}^{1/2}}{3}\eta^{3/2}+\frac{E_{\fed{r}}}{\tilde{F}^{1/2}}\eta^{1/2}\right]\right),\label{eq:f_eta}
\end{align}
$\Phi_\nu(\xi,\varphi)$ is the ionization channel function defined by
\begin{align}
 \left[\dpa{}{\xi} \xi\dpa{}{\xi} + \frac{1}{4\xi} \dpat{\varphi}+\tilde{Z}+\frac{E_{\fed{r}}\xi}{2}-\frac{\tilde{F}\xi^2}{4}  \right] \Phi_\nu(\xi,\varphi) & = \beta_\nu \Phi_\nu(\xi,\varphi),\label{eq:xi_phi_ad_eq}
\end{align}
and $\nu=(n_\xi,m)$ is a set of parabolic quantum numbers labeling the different ionization channels, see Fig.~\ref{fig:parab_coord}. With our choice of $F\geq 0$ the potential in Eq.~(\ref{eq:xi_phi_ad_eq}) goes to infinity as $\xi$ goes to infinity, so the parabolic channels $\nu$ are purely discrete.


The full wave function can be expressed as a linear combination, discrete in $\nu$, continuous in $k$, of the $f(\fed{r},k)g(R,k)$ products,
\begin{align}
  \Psi(\fed{r},R)|_{\eta\to\infty} & = \sum_\nu \int_0^\infty C_\nu(k)\eta^{-1/2}f(\eta)\Phi_\nu(\xi,\varphi) g(R,k) \frac{dk}{2\pi},\label{eq:wf_expansion}
\end{align}
where the asymptotic expansion coefficient $C_\nu(k)$ can be calculated by
\begin{align}
  C_\nu(k) &  = \frac{1}{ \eta^{-1/2}f(\eta)} \int_0^\infty g(R,k) \Braket{\Phi_{\nu}(\xi,\varphi)|\Psi(\fed{r},R)}_{(\xi,\varphi)} dR|_{\eta\to\infty}.\label{eq:spectrum_ampl_def}
\end{align}
$\Braket{{}\cdot{}}_{(\xi,\varphi)}$ indicates integration w.r.t. the coordinates $\xi$ and $\varphi$ over their full range. 
Note that the polar angle $\beta$, which we suppressed in the notation, only enters Eq.~(\ref{eq:spectrum_ampl_def}) through the wave function $\Psi(\fed{r},R)$.
The KER dissociation spectrum into the channel $\nu$ is defined in terms of these expansion coefficients by
\begin{align}
  P_\nu(k) & =\abs{C_\nu(k)}^2 .\label{eq:spectrum_def}
\end{align}
This is the main observable of interest. By inserting Eq.~(\ref{eq:f_eta}) and Eq.~(\ref{eq:spectrum_ampl_def}) and assuming $E_\fed{r}$ to be real, which is approximately the case for small $F$, we obtain
\begin{align}
  P_\nu(k) 
 & =\frac{\tilde{F}^{1/2}}{ 2\eta^{1/2} } \abs{\int g(R,k) \Braket{\Phi_{\nu}(\xi,\varphi)|\Psi(\fed{r},R)}_{(\xi,\varphi)} dR}^2 _{\eta\to\infty}.\label{eq:spec_expr}
\end{align}
The exact KER spectrum in the channel $\nu$ can thus be obtained by projecting the wave function on the channel state $\Phi_\nu(\xi,\varphi)$, and further projecting this on the continuum states $g(R,k)$ of the $U(R)$ potential. The total KER spectrum can then be obtained by summing over all the channels
\begin{align}
  P(k) & = \sum_\nu P_\nu(k).
\end{align}
In the $F\to 0$ limit the total rate can be obtained by
\begin{align}
  \Gamma & = \int_0^{\infty} P(k) \frac{dk}{2\pi}.
\end{align}

\subsection{Born-Oppenheimer approximation}
\label{sec:bo}
Now that we have a recipe for finding the exact KER spectrum, we consider some approximations for ease of predictions and gain in physical insight. We first consider the BO approximation, which appears in the limit $m_1=m_2\to\infty$. In this limit $m=1=q$, and the wave function takes the form $\Psi_\text{BO}(\fed{r},R)=\psi_e(\fed{r},R)\chi(R)$. The electronic and nuclear part of BO wave function fulfills the BO equations
\begin{subequations}
  \begin{align}
    \left[  - \frac{1}{2} \Delta_{\fed{r}}  + V(\fed{r}\Vcomma R)  +   Fz  - E_e(R;F)\right] \psi_e(\fed{r};R) & = 0,\label{eq:BO_elec_eq}\\
    \left[- \frac{1}{2M} \dd{R}   +  U(R)+E_e(R;F)   - E_{\text{BO}}(F)\right] \chi(R) & = 0.\label{eq:BO_nuc_eq}
  \end{align}
\end{subequations}
We impose zero boundary condition for the nuclear wave function
\begin{align}
  \chi(R=0) & = 0
\end{align}
and the following normalizations
\begin{align}
  \int_0^\infty dR\ {\chi^2(R)} & = 1,\\
  \int d^3\fed{r}\ \psi_e^2(\fed{r};R) & = 1.
\end{align}

In the asymptotic limit $\eta\to\infty$ the electronic Eq.~(\ref{eq:BO_elec_eq}) takes the same form as Eq.~(\ref{eq:as_x_eq}), and it can be written in parabolic coordinates in the same manner. The electronic wave function then takes the outgoing-wave form
\begin{align}
  \psi_e(\fed{r};R)|_{\eta\to\infty} & =  \eta^{-1/2}f(\eta) \sum_\nu f_\nu(R) \Phi_\nu(\xi,\varphi) ,
\end{align}
where $f(\eta)$ is from Eq.~(\ref{eq:f_eta}) and $\Phi_\nu(\xi,\varphi)$ are solutions of Eq.~(\ref{eq:xi_phi_ad_eq}), with $E_\fed{r}$ replaced by $E_e(R;F)$ in both. The asymptotic coefficient $f_\nu(R)$ defines the ionization amplitude in channel $\nu$~\cite{PhysRevA.84.053423}. The partial electronic ionization rate is given by
\begin{align}
  \Gamma_{e,\nu}(R) & = \abs{f_\nu(R)}^2\label{eq:partial_elec_rate}.
\end{align}
By considering the flux of the electron probability through a surface at large negative $z$, one can show that in the weak field limit the total electronic rate $\Gamma_e(R)=-2\imag(E_e(R;F))$ is given as a sum over $\nu$ of all the partial electronic rates.

By inserting the BO wave function into Eq.~(\ref{eq:spec_expr}) we obtain
\begin{align}
  P_\nu(k)
  & =  \abs{\int_0^\infty g(R,k) f_{\nu}(R)\chi(R)dR}^2 .\label{eq:Spec_BO}
\end{align}
The separation of electronic and nuclear coordinates in the BO approximation means that this expression for the KER spectrum does not contain any explicit reference to electronic coordinates, as opposed to the expression for the exact spectrum Eq.~(\ref{eq:spec_expr}). Equation~(\ref{eq:Spec_BO}) is similar to a result previously put forward in the literature [Eq.~(1) of Ref.~\cite{PhysRevLett.92.163004}], except that the correct complex ionization amplitude $f_\nu(R)$ was taken as $\sqrt{\Gamma_{e,\nu}(R)}$. In the cases we have considered, the phase variations of $f_\nu(R)$ are sufficiently small that they can be safely neglected, explaining the successful use of the aforementioned replacement in Ref.~\cite{PhysRevLett.92.163004}, but this is not generally true.

To evaluate the integral in Eq.~(\ref{eq:Spec_BO}) we will use the reflection principle~\cite{PhysRev.32.858,:/content/aip/journal/jcp/58/9/10.1063/1.1679721}. At the heart of the reflection principle lies an important mathematical component which we denote the reflection approximation~\cite{:/content/aip/journal/jcp/58/9/10.1063/1.1679721}. This approximation amounts to setting
\begin{align}
  g(R,k) 
         & = \sqrt{-2\pi\dpa{R_t}{k}}\delta\left(R-R_t\right),\label{eq:g_as_Delt}
\end{align}
which is exact in the $M\to\infty$ limit. In Eq.~(\ref{eq:g_as_Delt}) $R_t$ is the classical turning point~\footnote{We consider monotonically decreasing $U(R)$ potentials, so there is only one classical turning point.} for the $g(R,k)$ function defined by
\begin{align}
  {U}(R_t)=E_R=\frac{k^2}{2M}.
\end{align}
In order to determine the derivative $\dpa{R_t}{k}$ the form of the dissociative $U(R)$ potential must be known.
Inserting Eq.~(\ref{eq:g_as_Delt}) in Eq.~(\ref{eq:Spec_BO}) yields
\begin{align}
  P_\nu(k)
  & =  2\pi\abs{\dpa{R_t}{k}}\Gamma_{e,\nu}(R_t) \abs{\chi(R_t)}^2 .\label{eq:Spec_result_BO}
\end{align}
This result shows that using the reflection approximation in conjunction with the BO approximation we obtain a KER spectrum that is expressed as a product of a Jacobian factor, the electronic rate and the field-dressed nuclear wave function [Eq.~(\ref{eq:BO_nuc_eq})]. This is a lot simpler to calculate than evaluating either integrals in Eqs.~(\ref{eq:spec_expr}) or (\ref{eq:Spec_BO}), and is easily reversed to give a way to image the field-dressed nuclear wave function, and it is applicable to any molecule with a dissociative BO curve.


\subsection{Weak-field limit within the Born-Oppenheimer approximation}
\label{sec:elec_wfat}

The exact electronic rate $\Gamma_{e,\nu}(R)$ is often not available, since finding it requires solving the electronic problem Eq.~(\ref{eq:BO_elec_eq}), which is a highly non-trivial task for many systems. In such cases the weak-field asymptotic theory (WFAT) \cite{PhysRevA.89.013421,PhysRevA.84.053423,PhysRevA.87.043426} can be employed to obtain the rate. WFAT is an analytic theory which expresses the ionization rate in terms of properties of the field-free state. It is applicable in the weak-field limit.

Let $\beta^{(0)}_\nu(R)$ and  $\Phi_\nu^{(0)}(\xi,\varphi;R)$ denote the adiabatic eigenvalues and eigenfunctions solving Eq.~(\ref{eq:xi_phi_ad_eq}) for $F=0$ with the field-free electronic energy $E_e(R;F=0)$ replacing $E_\fed{r}$. Ref.~\cite{PhysRevA.84.053423} provides analytic expressions for these quantities. In terms of these the asymptotic field-free electronic wave function can be written
\begin{align}
  \psi_{e,0}(\fed{r};R)|_{\eta\to\infty}& = \sum_\nu  f_\nu(R,F=0)  \eta^{\beta_\nu^{(0)}(R)/\varkappa(R)-1/2}e^{-\varkappa(R)\eta/2}\Phi_\nu^{(0)}(\xi,\varphi;R),\label{eq:field_free_elec_wf}
\end{align}
where
\begin{align}
  \varkappa(R) & = \sqrt{-2E_e(R;F=0)}.
\end{align}
The electronic WFAT rate is then given by~\cite{PhysRevA.84.053423}
\begin{align}
  \Gamma_{e,\nu}^\text{WFAT}(R) & = \abs{f_\nu(R,F=0) }^2 W_\nu(R)\label{eq:rate_WFAT_elec}
\end{align}
where the field factor $W_\nu(R)$ is defined by
\begin{align}
W_\nu(R)= \frac{{\varkappa(R)}}{2}\left(\frac{4{\varkappa^2(R)}}{ F }\right)^{2\frac{\beta_\nu^{(0)}(R) }{{\varkappa(R)}}} \exp\left(-\frac{{2\varkappa^3(R)}}{3F } \right),\label{eq:field_factor_elec}
\end{align}
and the asymptotic coefficients $f_\nu(R,F=0)$ can be found from the electronic wave function by inversion of Eq.~(\ref{eq:field_free_elec_wf})
\begin{align}
 f_\nu(R,F=0) & = \left.\frac{  \Braket{\Phi_\nu^{(0)}(\xi,\varphi;R)|\psi_{e,0}(\fed{r};R)}_{(\xi,\varphi)}}{   \eta^{\beta_\nu^{(0)}(R)/\varkappa(R)-1/2}e^{-\varkappa(R)\eta/2}}\right|_{\eta\to\infty}.
\end{align}

\subsection{Weak-field asymptotic theory}
\label{sec:full_wfat}

WFAT can also be applied for the exact state, and not just in the BO approximation as above. In this section we will give the pertaining formulas. Let, as before, $\beta^{(0)}_\nu$ and  $\Phi_\nu^{(0)}(\xi,\varphi)$ denote the adiabatic eigenvalues and eigenfunctions solving Eq.~(\ref{eq:xi_phi_ad_eq}) for $F=0$ now with the field-free energy $E(F=0)$. In terms of these the asymptotic field-free wave function can be written~\cite{PhysRevA.84.053423}
\begin{align}
  \Psi_0(\fed{r},R)|_{\eta\to\infty}& = \sum_\nu \int_0^\infty C_\nu(k,F=0) g(R,k) \eta^{\beta_\nu^{(0)}/\varkappa(k)-1/2}e^{-\varkappa(k)\eta/2}\Phi_\nu^{(0)}(\xi,\varphi)\frac{dk}{2\pi}\label{eq:field_free_wf}
\end{align}
where
  \begin{align}
    \varkappa(k) 
                 & = \sqrt{2\left(\frac{1}{2M}k^2-E(F=0)\right)}.
  \end{align}
The WFAT \cite{PhysRevA.84.053423} yields the following expression for the KER spectrum
\begin{align}
  P_{\nu}^{\text{WFAT}}(k) & = \abs{C_\nu(k,F=0)}^2W_\nu(k),\label{eq:spec_WFAT}
\end{align}
where the field factor $W_\nu(k)$ is given by
\begin{align}
  W_\nu(k)= \frac{{\varkappa(k)}}{2}\left(\frac{4\sqrt{m}{\varkappa^2(k)}}{ qF }\right)^{2\frac{\beta_\nu^{(0)} }{{\varkappa(k)}}} \exp\left(-\frac{2{\sqrt{m}\varkappa^3(k)}}{3qF } \right),\label{eq:full_field_factor}
\end{align}
and the field-free asymptotic coefficients $C_\nu(k,F=0)$ can be found by inversion of Eq.~(\ref{eq:field_free_wf})
\begin{align}
 C_\nu(k,F=0)  & = \left.\frac{ \int_0^\infty g(R,k)\Braket{\Phi_\nu^{(0)}(\xi,\varphi)| \Psi_0(\fed{r},R)}_{(\xi,\varphi)} dR}{ \eta^{\beta_\nu^{(0)}/\varkappa(k)-1/2}e^{-\varkappa(k)\eta/2}}\right|_{\eta\to\infty}.
\end{align}


\section{Illustrative 1D calculations}
\label{sec:1d-calculation}
Solving Eq.~(\ref{eq:schrodinger}) in 3D is a computationally heavy task, so we have used a 1D model to illustrate our central points. In this section we compare exactly calculated KER spectra with those obtained through the BO approximation, Eq.~(\ref{eq:Spec_result_BO}), and the WFAT, Eq.~(\ref{eq:spec_WFAT}), within this 1D model. In the following we will consider a model of H$_2^+$ as an example. The potentials we consider are thus
\begin{subequations}
  \begin{align}
    U(R)& = \frac{1}{R},\\
    V(\elecvar\Vcomma R)& =  -\sum_{\pm}\frac{1}{\sqrt{\left(\elecvar\pm\frac{R}{2}\right)^2+a(R)}},
  \end{align}
\end{subequations}
with $m_1=m_2=1836$ and $q_1=q_2=1$. The interaction between the nuclei and the electrons $V(\elecvar\Vcomma R)$ is described by a soft-core Coulomb potential. The function $a(R)$ is chosen in such a way that the BO potential of this potential reproduces the BO potential energy curve of 3D H$_2^+$ \cite{PhysRevA.53.2562,PhysRevA.67.043405,PhysRevLett.98.253003}. We use the method described in Ref.~\cite{PhysRevA.91.013408} to solve the 1D equivalent of Eq.~(\ref{eq:schrodinger}) 
 given by
\begin{align}
 \left[- \frac{1}{2M} \dpat{R}  - \frac{1}{2m} \dpat{z} +  U(R) + V\left(z\Vcomma R\right)  +   \qx zF - E(F)\right] \Psi(z,R) & = 0.\label{eq:schrodinger_1D}
\end{align}
In the 1D model the index $\nu$, which describes what happens in the paraboloids of constant $\eta$ 'transversal' to $z$, is of no meaning, and it hence does not appear in any of the 1D equivalents of the 3D equations. The 1D equivalent of the exact KER spectrum Eq.~(\ref{eq:spec_expr}) is
\begin{align}
  P(k) 
 & =\frac{(2Fq\abs{z})^{1/2}}{ m^{1/2} } \abs{\int g(R,k) \Psi(z,R) dR}^2_{z\to -\infty} .\label{eq:spec_expr_1D}
\end{align}
Equations~(\ref{eq:Spec_BO}) and (\ref{eq:Spec_result_BO}) apply to the 1D case with appropriately redefined quantities. The WFAT expressions Eqs.~(\ref{eq:rate_WFAT_elec}), (\ref{eq:field_factor_elec}) and (\ref{eq:spec_WFAT}), (\ref{eq:full_field_factor}) are the same as in the 1D case, but the asymptotic coefficients are now found from
\begin{align}
  f(R)&=\left.\frac{ \psi_{e,0}(z;R)}{\abs{z}^{Z/\varkappa(R)}e^{-\varkappa(R)\abs{z}}}\right|_{z\to -\infty},\label{eq:1D_asymp_coeff}
\end{align}
and
\begin{align}
  C(k)&=\left.\frac{\int_0^\infty g(R,k) \Psi_{0}(z,R)dR}{\abs{z}^{mZ/\varkappa(k)}e^{-\varkappa(k)\abs{z}}}\right|_{z\to -\infty}.
\end{align}
In Eq.~(\ref{eq:1D_asymp_coeff}), $\psi_{e,0}(z;R)$ denotes the field-free 1D electronic BO wave function.
\subsection{From wave function to KER spectrum}

\begin{figure}
  \centering
  \includegraphics{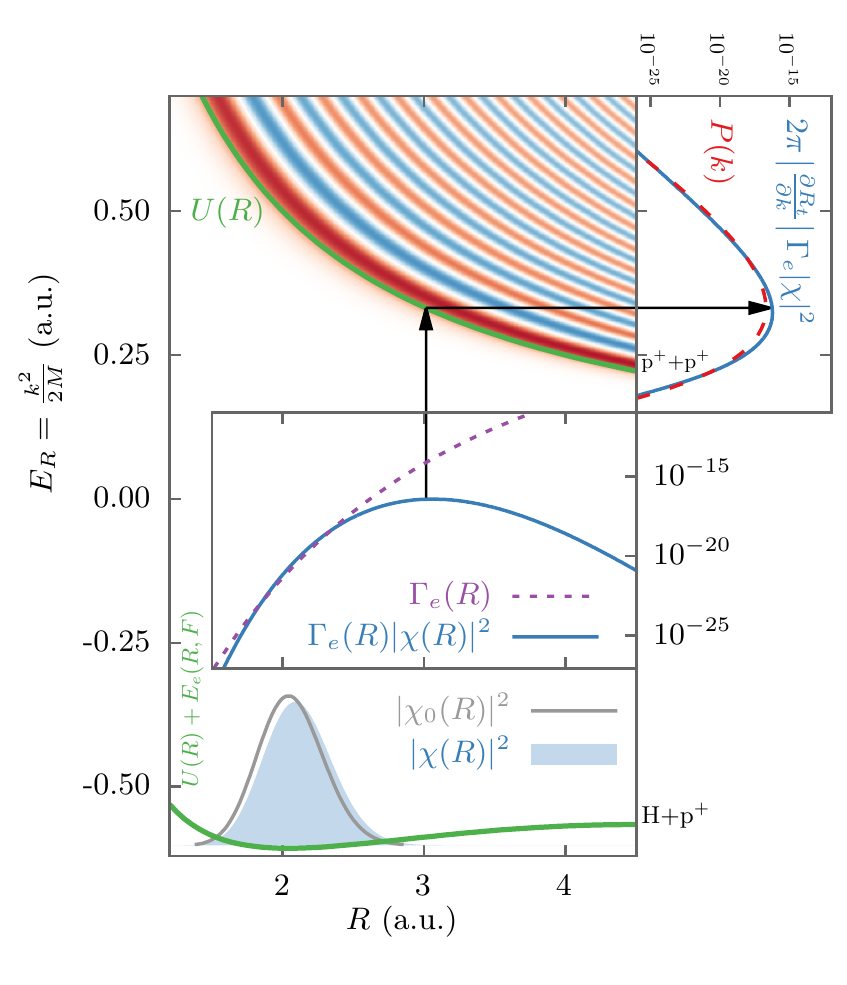}
  \caption{(Color online) The field dressed nuclear wave function $\abs{\chi(R)}^2$ (light blue shaded area in the lower $U(R)+E_e(R;F)$ BO curve) is multiplied by the electronic rate $\Gamma_e(R)$ (dashed purple line) and reflected in the dissociative $U(R)$ BO curve to give a KER spectrum (solid blue line in upper right corner, [Eq.~(\ref{eq:Spec_result_BO})]), using the relation $U(R)=\frac{k^2}{2M}$ to translate $k$ into $R$. This is compared to the exact KER spectrum $P(k)$ (red dashed line, [Eq.~(\ref{eq:spec_expr_1D})]). A field strength of $F=0.034$ was used for this calculation. The solid gray line in the lower part of the figure shows the field-free nuclear wave function $\abs{\chi_0(R)}^2$. The surface plot in the upper part of the figure shows the continuum states $g(R,k)$ of the $U(R)$ potential, these are solutions of Eq.~(\ref{eq:as_R_eq}).}
  \label{fig:spec_1D_ground}
\end{figure}

Figure~\ref{fig:spec_1D_ground} illustrates how the BO approximation can be used in conjunction with the reflection principle to determine the KER spectrum. The figure shows a calculation for the ground state of the H$_2^+$ model at $F=0.034$. The field dressed nuclear wave function $\abs{\chi(R)}^2$ is multiplied by the electronic rate $\Gamma_e(R)$. The exponential dependence of the electronic rate $\Gamma_e(R)$ on the internuclear coordinate means that the product $\Gamma_e(R)\abs{\chi(R)}^2$ (see [Eq.~(\ref{eq:Spec_result_BO})]) has its maximum at a value of $R\approx 3$, which is significantly different from the maximum of the bare nuclear wave function at $R_0\approx 2$. This in turn means that the transition to the continuum which is determined by the product $\Gamma_e(R)\abs{\chi(R)}^2$ and not the bare nuclear wave function is far from 'vertical' in $R$ with respect to the initial nuclear wave function, and the spectrum peaks at a lower energy around $1/R\approx 0.33$ and not at $1/R_0\approx 0.5$.

Using WFAT within the BO approximation we can make a statement about in which direction the maximum of the spectrum shifts when the field is varied. In these approximations the main dependence of the electronic rate on the field is contained in the exponent $-\frac{2\varkappa^3(R)}{3F}$, see Eq.~(\ref{eq:field_factor_elec}). The electronic energy $E_e(R;F=0)$, in terms of which $\varkappa(R)$ is defined, generally depends very much on the system considered. In the case of H$_2^+$ it is a monotonically increasing function of $R$, since when the two potential wells around each of the nuclei start to overlap the electron is more tightly bound. This in turn means that the electronic rate is an increasing function of $R$, as can also be seen in Fig.~\ref{fig:spec_1D_ground}. When the strength of the field increases the exponent $-\frac{2\varkappa^3(R)}{3F}$ grows, but at the same time the slope of this exponent with respect to $R$ decreases, since $\varkappa^3(R)$ is multiplied by a smaller number. The smaller slope means that the location of the maximum of the product $\Gamma_e(R)\abs{\chi(R)}^2$ is shifted less from the maximum of $\abs{\chi(R)}^2$ as the field strength increases, and conversely, as the field strength is decreased the maximum of the product $\Gamma_e(R)\abs{\chi(R)}^2$ is shifted more towards larger $R$. These shifts are directly reflected in the spectrum, which is given as the reflection of the $\Gamma_e(R)\abs{\chi(R)}^2$ product in the BO and reflection approximations.

\begin{figure}
  \centering
  \includegraphics{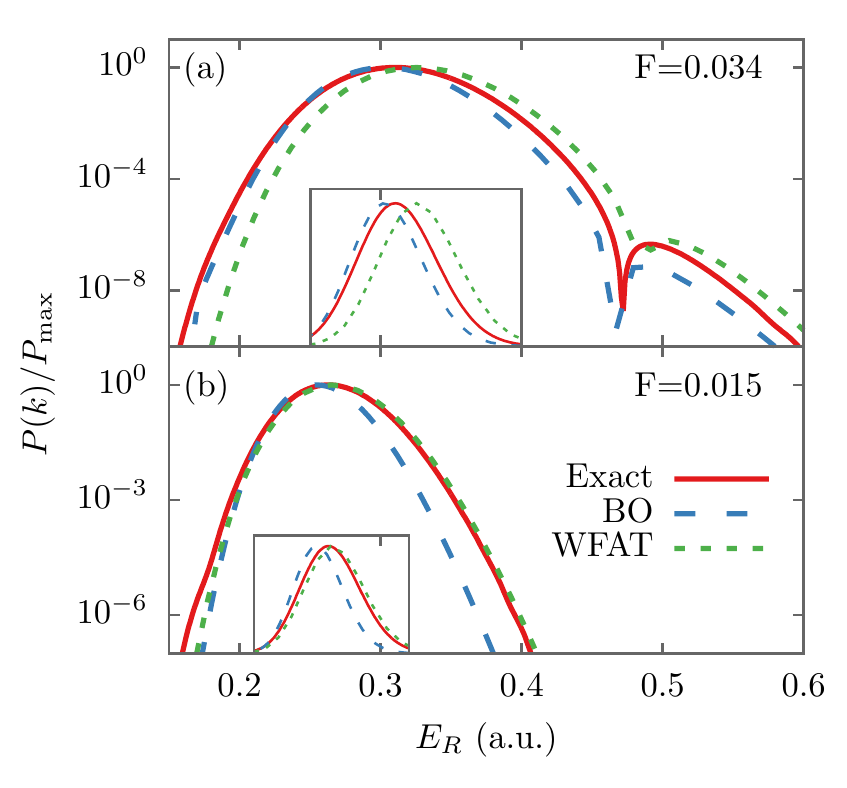}
  \caption{(Color online) KER Spectra normalized to their maxima. Solid (red) line: $P(k)$ [Eq.~(\ref{eq:spec_expr_1D})]. Dashed dotted (blue) line: BO combined with reflection principle [Eq.~(\ref{eq:Spec_result_BO})]. Short dashed (green) line: WFAT [1D equivalent of Eq.~(\ref{eq:spec_WFAT})]. The insets show the normalized KER spectra on a linear scale. The critical field for use of BO [Eq.~(\ref{eq:F_BO})] is for H$_2^+$: $F_\text{BO}=0.0315$. (a) $P_\text{max}^\text{BO}/P_\text{max}^\text{Exact}=2.32$ and $P_\text{max}^\text{WFAT}/P_\text{max}^\text{Exact}=0.29$. (b) $P_\text{max}^\text{BO}/P_\text{max}^\text{Exact}=95.0$ and $P_\text{max}^\text{WFAT}/P_\text{max}^\text{Exact}=0.54$. }
  \label{fig:spec_1D}
\end{figure}

Figure~\ref{fig:spec_1D} shows KER spectra obtained using as initial state the first vibrationally exited state of H$_2^+$. We have chosen to show these results as they are for the lowest state with a non-trivial nodal structure in $R$. In the figure two different field strengths are considered. In the top panel we see that the nodal structure of the nuclear wave function is reflected in the KER spectrum, although one peak is a lot larger than the other. This asymmetry can be understood in the BO approximation, see Eq.~(\ref{eq:Spec_result_BO}), as due to the fact that the electronic rate $\Gamma_e(R)$ has an exponential dependence on $R$. In the WFAT it can be understood as resulting from the exponential dependence of the field factor [Eq.~(\ref{eq:full_field_factor})] on $k$. For the lower field strength the structures at $E_R>0.4$ are not visible as the KER spectrum falls below the numerical precision limit of our calculation.

For the large field strength [Fig.~\ref{fig:spec_1D}(a)] we see that the BO KER spectrum has a shape much closer to the exact KER spectrum than for the lower field strength. Also the maximum value of the BO KER spectrum is more than an order of magnitude closer to the maximum value of the exact KER spectrum for the larger field strength. This can be understood on the basis of the retardation argument provided in Ref.~\cite{rtd}: The BO approximation is expected to hold as long as the electron is close enough to the nuclei that the time it takes for the electron to go to its present location from the nuclei is shorter than the time it takes for the nuclei to move. 
A typical electron velocity can be estimated as $\varkappa_e=\sqrt{-2E_e(R_0;0)}$, where $R_0$ is the equilibrium internuclear distance, which for H$_2^+$ is $R_0=2$. A typical time scale for the nuclear motion can be estimated as $T=\frac{1}{2\omega_e}$, where $\omega_e$ is obtained by expanding the BO potential around $R_0$ to second order $U(R)+E_e(R;0)\approx U(R_0)+E_e(R_0;0)+\frac{1}{2}M\omega_e^2(R-R_0)^2$. Using these estimates Ref.~\cite{rtd} defines a critical distance 
\begin{align}
  \elecvar_\text{BO} & = \varkappa_e T=\frac{\varkappa_e}{2\omega_e},\label{eq:z_BO}
\end{align}
such that for $\abs{\elecvar}<\elecvar_\text{BO}$ we expect BO to work well, while for $\abs{\elecvar}>\elecvar_\text{BO}$ we expect it to break down. Since the magnitude of the wave function is essentially unchanged after the tunneling, the BO approximation is expected to work well when the outer turning point is within this $\elecvar_\text{BO}$ distance, so a critical field 
\begin{align}
  F_\text{BO} & = 2 \varkappa_e\omega_e \label{eq:F_BO}
\end{align}
can be estimated, such that the BO approximation is expected to give good results for larger fields, but fail for smaller fields. The two field strengths of Fig.~\ref{fig:spec_1D} lies on either side of this critical field, which for the system under consideration is $F_\text{BO}=0.0315$. As we increase the field strength further the BO gives even better results.

For the lower field strength where BO fails we can apply the WFAT, see Sec.~\ref{sec:full_wfat}. In Fig.~\ref{fig:spec_1D} we see that the shape of the WFAT KER spectrum indeed is closer to the exact KER spectrum than the BO KER spectrum for the weaker field strength, and it is also closer in magnitude to the maximum value. For the larger field strength the WFAT KER spectrum is further from the exact KER spectrum in both shape and magnitude.


\subsection{From KER spectrum back to wave function}
\begin{figure}
  \centering
  \includegraphics{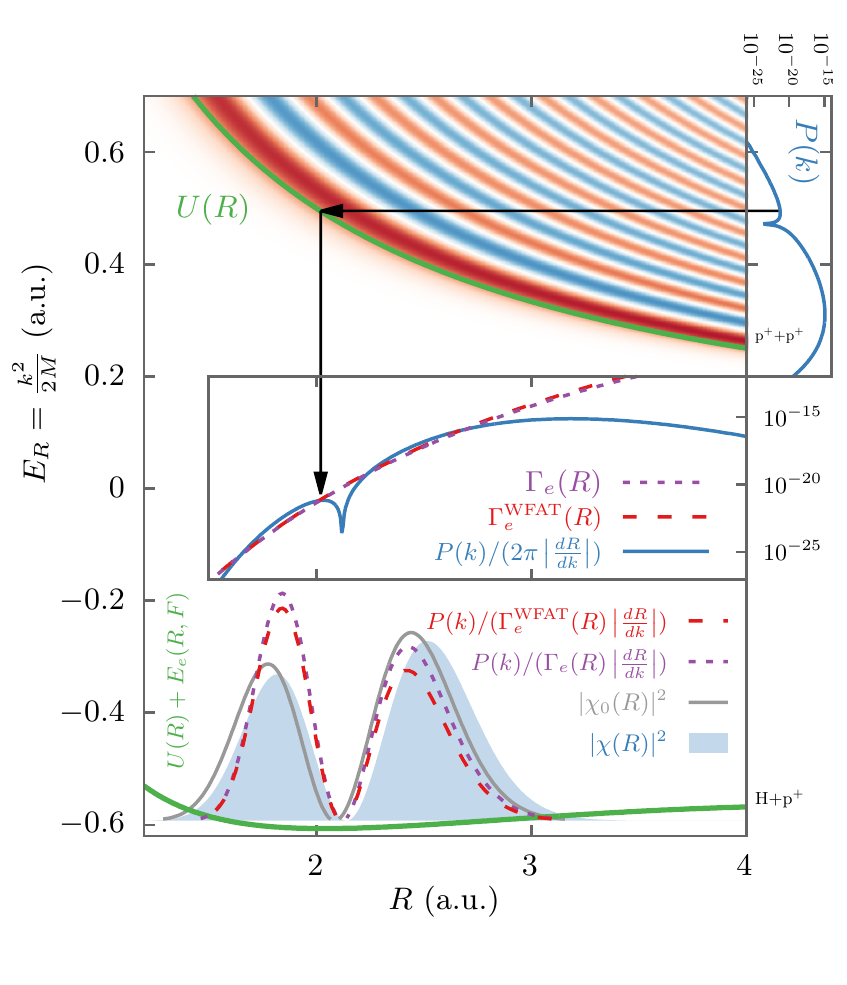}
  \caption{(Color online) From the exact KER spectrum $P(k)$ (upper right corner, [Eq.~(\ref{eq:spec_expr_1D})]) at $F=0.034$ the magnitude of the asymptotic wave function has been found by reversing the reflection principle, giving $P(k)/(2\pi\abs{\frac{dR}{dk}})$, using the relation $U(R)=\frac{k^2}{2M}$ to translate $k$ into $R$. From this, the field-dressed nuclear wave function has been imaged by dividing with the electronic rate $\Gamma_e(R)$ and normalizing. In the lowest part of the plot, the short dashed (purple) line shows this imaging using the exact electronic rate $\Gamma_e(R)=-2\imag(E_e(R;F))$, the long dashed (red) line shows it using the BO WFAT approximation $\Gamma_e^\text{WFAT}(R)$ [Eq.~(\ref{eq:rate_WFAT_elec})]. The solid gray line shows the field-free nuclear wave function $\abs{\chi_0(R)}^2$. The shaded (light blue) area shows the field-dressed nuclear wave function $\abs{\chi(R)}^2$. The surface plot in the upper part of the figure shows the continuum states $g(R,k)$.}
  \label{fig:Reconstructed_chi}
\end{figure}

The field dressed nuclear wave function can be imaged from a measurement of the KER spectrum by inverting Eq.~(\ref{eq:Spec_result_BO}) for fields sufficiently large that the BO approximation applies. To demonstrate this we have taken the exact KER spectrum from our calculation at $F=0.034$ for the first vibrationally exited state and divided it by the Jacobian factor and the electronic rate to obtain an image of the nuclear density. Since an experimental KER spectrum is typically not known on an absolute scale, we have then normalized this quantity. In a calculation on a more complicated system than the one considered here the exact electronic rate is often not available, so we also show the result using the WFAT approximation for the electronic rate [Eq.~(\ref{eq:rate_WFAT_elec})]. The results are compared to the nuclear wave function known from the calculation in our model in Fig.~\ref{fig:Reconstructed_chi}. They do not agree perfectly, but the nodal structure is correctly reproduced.

\begin{figure*}
  \centering
  \includegraphics{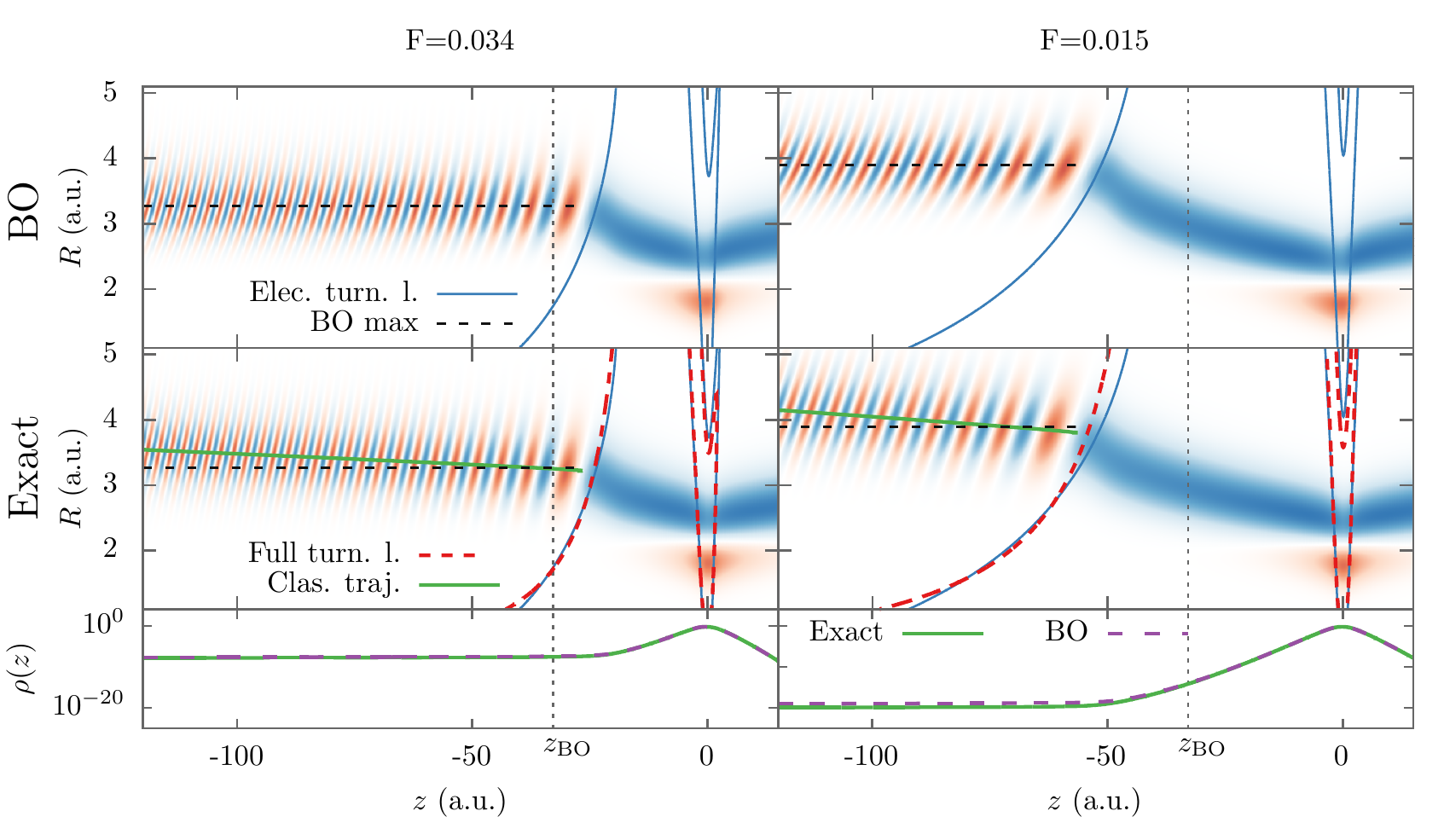}
  \caption{(Color online) Real part of wave function normalized with the electron density $\rho(\elecvar)=\int \abs{\Psi(\elecvar,R)}^2 dR$ for two different fields strengths. Solid (blue) lines: Electronic turning lines $\real (E_e(R;F))=V(\elecvar\Vcomma R)+Fq\elecvar$. Dashed (red) lines: Full turning lines $\real(E(F))=V(\elecvar\Vcomma R)+U(R)+Fq\elecvar$. In the upper panels the dashed black line shows a constant line that coincides with the maximum of the BO wave function. In the middle panels the green line additionally shows a classical trajectory that coincides with the maximum of the exact wave function. The critical BO distance [Eq.~(\ref{eq:z_BO})] is shown with a vertical dashed line at $-\elecvar_\text{BO}=-32.9$.}
  \label{fig:wf}
\end{figure*}

For smaller field strengths where the BO is not applicable this type of imaging is not possible. The KER spectrum, however, does give us access to the asymptotic wave function, as it is the norm square of the expansion coefficients of this, see Eq.~(\ref{eq:wf_expansion}). For the cases we have looked at, the phase of the asymptotic coefficient $C(k)$ varies very little over the range where it has support. In our model we have access to the full wave function, and this we show in Fig.~\ref{fig:wf}. The imaging through the 1D equivalent of Eq.~(\ref{eq:wf_expansion}) would only give access to the part at large negative $\elecvar$.

In the classically allowed region at large negative $\elecvar$ the maximum of the wave function follows a classical trajectory. This is a prediction of the WKB theory, which applies as long we are not too close to the turning line. The classical trajectories can be found using Newton's second law
\begin{subequations}
  \begin{align}
    m \ddot{z} +\dpa{}{z}V(z;R) + Fq & = 0,\\
    M \ddot{R} +\dpa{}{R}V(z;R) + \dpa{}{R}U(R)  & = 0.
  \end{align}\label{eq:clas_traj_newton}
\end{subequations}
A tempting choice of initial condition for the differential Eqs.~(\ref{eq:clas_traj_newton}) would be to choose the $(\elecvar,R)$ values at the intersection of the outer turning line and the maximum ridge of the wave function, with zero velocity in both $\elecvar$ and $R$ direction. However, the WKB fails near the turning line, and therefore we cannot expect the wave function to follow a classical trajectory here. Instead we have chosen as initial condition some point at the maximum of the wave function at a large negative $\elecvar$ value away from the turning line. The influence of the $V(z;R)$ potential can be neglected for sufficiently large negative $\elecvar$, in this region we can write the separated energy conservation equations
\begin{subequations}
  \begin{align}
    \frac{1}{2}m \dot{\elecvar}^2 + Fq\elecvar & = E-\frac{1}{2M}k^2,\\
    \frac{1}{2}M \dot{R}^2 + {U}(R) & = \frac{1}{2M}k^2.
  \end{align}\label{eq:clas_traj_energy}
\end{subequations}
The initial velocities have then been determined from Eqs.~(\ref{eq:clas_traj_energy}), using the real part of the total (quantum) energy for $E$ and the $k$ at which the KER spectrum $P(k)$ [Eq.~(\ref{eq:spec_expr_1D})] peaks. The classical trajectories shown in Fig.~\ref{fig:wf} were found using such initial conditions, and then propagated inwards.

From Fig.~\ref{fig:wf} it can be seen that contrary to the exact wave function, the position of the ridge of the BO wave function in $R$ does not change with $\elecvar$. This is expected as the BO approximation appears in the limit of infinite nuclear mass, so classical motion in the nuclear coordinate is not possible. The asymptotic wave function that we can image using Eq.~(\ref{eq:wf_expansion}) is therefore a non-BO wave function.

It might seem strange that the BO is able to give the correct KER spectrum when the spectrum is the norm square of the expansion coefficients of the asymptotic wave function, and the BO gives a wrong description of this asymptotic wave function. However, the fact that the BO wave function does not obtain a probability current (or velocity in the classical picture) in the $R$-direction does not alter its projection on the continuum states. The important point is whether the BO wave function is similar to the exact wave function as it emerges at the outer turning line after tunneling, and this is the case if the turning line is within the critical BO distance $\elecvar_\text{BO}$ [Eq.~(\ref{eq:z_BO})].

In Fig.~\ref{fig:wf} we also see that for the larger field strength the tunneling is completed before the critical BO distance is reached, contrary to at the smaller field strength. We see that for the large field strength the electronic and full turning lines agree quite well in the region where most of the wave function is localized, but for the smaller field strength they do not.

\begin{figure}
  \centering
  \includegraphics{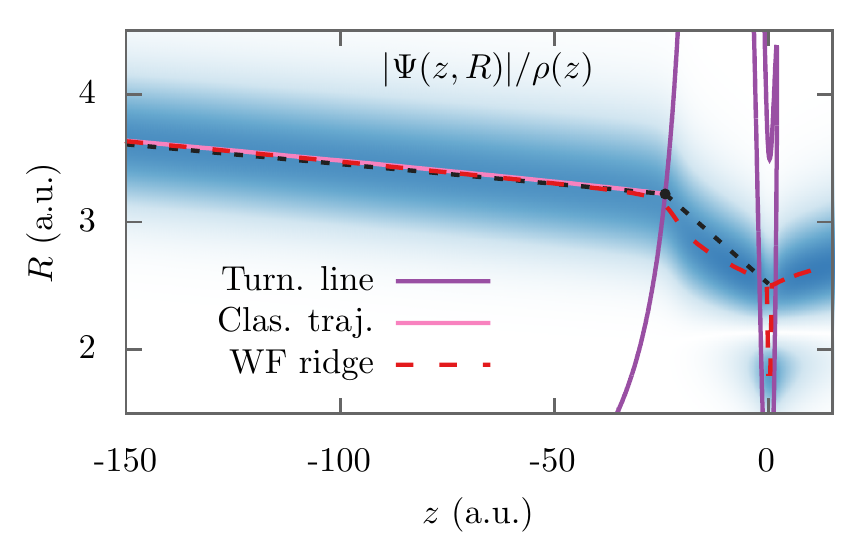}
  \caption{(Color online) Absolute value of wave function normalized with the electron density $\rho(\elecvar)=\int \abs{\Psi(\elecvar,R)}^2 dR$ for $F=0.035$. Solid purple lines: Full turning lines $\real(E(F))=V(\elecvar\Vcomma R)+U(R)+Fq\elecvar$. The long dashed red line shows for each $z$ the $R$ at which the wave function $\abs{\Psi(z,R)}$ has its maximum. The solid pink line shows a classical trajectory [Eq.~(\ref{eq:clas_traj_newton})]. The black dot at the end of the classical trajectory is the exit point $(z_{k_\text{max}},R_{k_\text{max}})$ determined from the maximum of the spectrum $k_\text{max}$ (see main text). The short dashed lines are the simple straight line estimates for the tunneling and initial classical motion described around Eq.~(\ref{eq:directions}).}
  \label{fig:wf_max}
\end{figure}

\subsection{A refraction phenomenon}

One can notice that a phenomenon reminiscent of light refraction occurs for the wave function around the turning line in Fig.~\ref{fig:wf}. It is evident, that the direction in which the maximum of the wave function 'moves' changes noticeably at the  turning line, when the wave function escapes from the classically forbidden tunneling region into the classically allowed region. The change of direction is due to the two different types of 'motion' involved. When the wave function emerges from the tunneling region it has essentially zero average velocity in the $R$ direction. This means that we can apply the reflection principle in reverse on the spectrum to find the $R_{k_\text{max}}$ coordinate at which the maximum of the wave function emerges from the tunneling region by the relation $U(R_{k_\text{max}})=\frac{k_\text{max}^2}{2M}$, where $k_\text{max}$ is the value of $k$ for which the spectrum $P(k)$ has its maximum. The $z$ value corresponding to this $R_{k_\text{max}}$ can then by found by considering the turning line $V(z_{k_\text{max}},R_{k_\text{max}})+U(R_{k_\text{max}})+Fqz_{k_\text{max}}=\real E$.

In Fig.~\ref{fig:wf_max} we see that near the turning line the location of the wave function ridge differs from the classical trajectory. This is expected, since the prediction that the wave function ridge should follow a classical trajectory comes from WKB theory, which fails near the turning line. Nevertheless, we can roughly describe the dissociative tunneling ionization process in two steps. First the system tunnels from the central region around $z=0$ to the exit point $(z_{k_\text{max}},R_{k_\text{max}})$. This motion can roughly be described by a straight line from the maximum of the nuclear wave function $\abs{\chi(R)}^2$ that has the largest $R$ value, since this is the maximum that will dominate the tunneling, to the exit point. Notice that this tunneling is not simply the electron tunneling out, but a correlated process involving both the electronic and nuclear degrees of freedom. In the classically allowed region the initial direction of the wave function from the exit point can be found from the classical trajectory: The initial slope of the classical trajectory that starts at the exit point $(z_{k_\text{max}},R_{k_\text{max}})$ with zero velocity in both $z$ and $R$ directions can be found to be 
\begin{align}
\left.\frac{\dot{z}}{\dot{R}}\right|_{\text{exit}}=\left.\frac{M}{m}\frac{\dpa{}{z}V(z;R) + Fq}{\dpa{}{R}V(z;R)+\dpa{}{R}U(R)}\right|_{(z_{k_\text{max}},R_{k_\text{max}})}.\label{eq:directions}
\end{align}
This is not exactly the trajectory that describes the motion of the wave function ridge, but it is quite close. These two directions are different as they come from different types of motion, and hence we see the refraction-like phenomenon at the turning line.

\section{Conclusion}
\label{sec:conclusion-outlook}

We have formulated theory for the dissociative tunneling ionization process, and derived exact formulas for the KER spectrum, as well as approximations in the framework of the BO and reflection approximations. We have demonstrated that the reflection principle can be used in conjunction with the BO approximation to image the field-dressed nuclear wave function from the KER spectrum. For weaker fields, where the BO approximation fails, the WFAT can be used to find the KER spectrum. We have also demonstrated a qualitative difference between asymptotic BO and exact wave functions, as the latter shows classical motion in the nuclear coordinate, whereas the former does not move at all due to the infinite nuclear mass of the BO approximation. Around the turning line the wave function exhibits a behavior similar to refraction of light.


\begin{acknowledgments}
This work was supported by the ERC-StG (Project No. 277767-TDMET), and the VKR center of excellence, QUSCOPE. The numerical results presented in this work were performed at the Centre for Scientific Computing, Aarhus \url{http://phys.au.dk/forskning/cscaa/}. O.~I.~T. acknowledges support from the Ministry of Education and Science of Russia (State Assignment No. 3.679.2014/K).
\end{acknowledgments}

%


\end{document}